\documentclass[12pt]{article}

\usepackage{amsbsy}

\begin{document}
\begin{center}
{\Large \bf
Bargmann-Michel-Telegdi equation and one-particle relativistic approach
\\}

\bigskip
{\large A. Della Selva}\\
\medskip
{\it Dipartimento di Scienze Fisiche, Universit\`a di Napoli\\
Istituto Nazionale di Fisica Nucleare, Sezione di Napoli\\
80125 Napoli, Italy}

\bigskip
{\large J. Magnin, L. Masperi}\\
\medskip
{\it Centro At\'omico Bariloche and Instituto Balseiro\\
Comisi\'on Nacional de Energ\'{\i}a At\'omica\\
and Universidad Nacional de Cuyo\\
8400 Bariloche, Argentina}

\end{center}

\begin{abstract}
\label{Abstract}
A reexamination of the semiclassical approach of the relativistic electron
indicates a possible variation of its helicity for electric and magnetic
static fields applied along its global motion due to zitterbewegung effects,
proportional to the anomalous part of the magnetic moment.
\end{abstract}
\section{Introduction}
\label{Introduction}

The one-particle interpretation of Dirac equation is an old problem which has
not a unique solution.  The usual Foldy-Wouthuysen method~\cite{Foldy} is
suitable for atomic physics where the electron velocity is
small but cannot be applied in presence of intense electromagnetic fields.
On the other hand the semiclassical limit~\cite{Rubinov} of Dirac equation
when the particle wavelength is small compared to the characteristic distance
of the electromagnetic potential leads to a trajectory determined by the
Lorentz force plus a spin precession given by the Bargmann-Michel-Telegdi
equation~\cite{Bargmann} (BMT).  The latter approximation is better for
relativistic velocities~\cite{Baier1} and might be applied also for particles
in laser fields~\cite{Chakrabarti}.

The purpose of the present work is to allow the influence in the semiclassical
approximation of the motion of the electron inside the scale of
the Compton length and the perturbation caused by the zero-point radiation
field.

Whereas non-relativistic quantum mechanics has equations of motion for
operators equivalent to the classical Hamilton ones, Dirac equation leads to
the picture that mass and charge do not coincide.  This has suggested
classical models~\cite{Barut} which double the number of variables and
correspond to equations of motion equivalent to the operatorial Dirac
ones, giving the spin as a result of zitterbewegung.  This analogy indicates
that zitterbewegung is twofold:  that of mass and a broader one of charge.
On the other hand this classical model does not give the average of
velocities which emerges from the semiclassical approximation of the
operatorial Dirac equations.  As a consequence we refer to BMT which involves
the average of only one velocity, taking it as the effective common
charge-mass one.

We have therefore reinterpreted BMT in such a way that, keeping its feature
of affecting ``Thomas $\frac{1}{2}$'' in the spin orbit interaction only
to Dirac magnetic moment but not to its anomalous
correction~\cite{Berestetskii}, allows a semiclassical explanation of the
latter.
This is so because, instead of the view that the fluctuations of the
zero-point radiation field can only decrease the effective
spin~\cite{Itzykson}, the velocity contribution to spin in BMT increases by
this fluctuation added to the zitterbewegung one~\cite{Masperi}.

The main result of the present paper is that the zero component of the
Pauli-Lubansky polarization vector defined as average of the instantaneous
variables may not vanish in the electron global rest frame because of the
zitterbewegung.  This is suggested both by the semiclassical
limit~\cite{Rafanelli} of the Dirac equation and its classical
analogy~\cite{Barut}.

A consequence of this would be the appearance of an additional term to
the spin-orbit interaction, extremely tiny even for very strong magnetic
fields applied to atoms, proportional to the anomalous pact of magnetic
moment.  But this consequence may be doubtful because of the low velocity
of the electron in an atom.  Another effect, also proportional to the
radiative correction to magnetic moment. which might be more trustful
corresponds to a small variation with time of the helicity of a free
electron to which strong longitudinal electric and magnetic fields are
applied.  This effect resembles the semiclassical interpretation of
chiral anomaly using its equivalence to the explicit symmetry breaking
for a very massive particle ~\cite{Gribov} in a model in which the
particle spin is entirely due to precession around a magnetic field
{}~\cite{Teryaev1}.  The relation of the anomaly effect on the electron
propagation with its anomalous moment correction has also been noticed
at the level of Feynman diagrams ~\cite{Teryaev2}.  It must be remarked
that our effect due to longitudinal fields is different from the
transverse polarization by radiation of electrons in storage rings
{}~\cite{Baier2}.

\section{Reinterpretation of BMT}
 \label{Reinterpretation of BMT}

We remind that normal BMT is

\begin{equation}
\frac{da^{\mu}}{d\tau} = c_{1} F^{\mu\nu}a_\nu + c_2 \frac{\pi^\mu}{m}
\frac{\pi_\nu}{m} F^{\nu\lambda}a_{\lambda}
\end{equation}

\noindent where $\pi^{\nu}$ is the kinetic momentum and $a^{\mu}$ the
Pauli-Lubansky
vector

\begin{equation}
a^{\mu} = \epsilon^{\mu\nu\rho\sigma} \frac{\pi_{\nu}}{m} M_{\rho\sigma}
\hspace{.3in}.
\end{equation}

By definition $\pi \cdot a = 0$ and in the rest frame $a^{\mu}$ reduces to a
3-vector  $\boldsymbol{\xi}$ which is interpreted as twice the average of spin.
$\tau$ is the proper time and $m$ the rest mass.

{}From eq.1 in rest frame for spatial indices $c_{1}=2{\mu}$, where ${\mu}$
is the magnetic moment.  Constancy of $\pi \cdot a=0$ in any frame together
with Lorentz force

\begin{equation}
\frac{d\pi^{\nu}}{d\tau} = eF^{\mu\nu} \frac{\pi_{\nu}}{m}
\end{equation}

\noindent leads to $-c_{2} = 2{\mu'} = 2(\mu - \frac{e}{2m})$.  In rest frame
$\frac{d}{d\tau}({\boldsymbol{\xi}}^2) = 0$ and, in general,
$\frac{d}{d\tau}(a^{2})=0$.

Our interpretation of BMT is based on considering it by covariance

\begin{equation}
\frac{da^{\mu}}{d\tau} = \tilde{c}_{1}F^{\mu\nu}a_{\nu} +
\tilde{c}_{2}\langle{u^{\mu}u_{\nu}}\rangle F^{\nu\lambda}a_{\lambda} +
\tilde{c}_{3} \frac{\pi^{\mu}}{m} \frac{\pi_{\nu}}{m}F^{\nu\lambda}a_{\lambda}
\end{equation}

\noindent where the second term contains the quadratic average of common
charge and mass instantaneous velocity $u^{\mu}$ different from the global
$\pi^{\mu}/m$.

In rest frame for only magnetic field just the first two terms of eq.4
contribute to the evolution of $\boldsymbol{\xi}$.  Since we assume that, in
absence of radiative  corrections, the second term represents the common
charge-mass zitterbewegung it must give one Bohr magneton corresponding to
the ratio of magnetic moment to angular momentum due to proportional density
of charge and mass.  Consequently, the first term represents the effect of
zitterbewegung of charge relative to mass and must account for the other
Bohr magneton of Dirac spin.  Therefore

\begin{equation}
\tilde{c}_{1} = \frac{e}{2m}\hspace{.4in},\hspace{.4in}
-\tilde{c}_{2}\frac{1}{3}\langle{\bf{u}}^2\rangle_{Z} = \frac{e}{2m}
\hspace{.4in}.\end{equation}

The additional radiative correction gives the right first-order anomalous
magnetic moment if we choose the zitterbewegung average

\begin{equation}
\langle{\bf{u}}^2\rangle_Z = 1 = \langle
{\frac{{\bf{v}}^2}{1-{\bf{v}}^2}} \rangle\hspace{.4in}.  \end{equation}

In fact, in rest frame, the ordinary semiclassical treatment of zero-point
radiation effect~\cite{Bjorken} for wavelengths larger than the Compton one
produces, in addition to the above average of ${\bf{u}}^2$ another

\begin{equation}
\langle{{\bf{u}}^2}\rangle_{R} = \frac{\alpha}{\pi}
\end{equation}

\noindent considering that only quadratic variables are affected by
fluctuations.
To obtain eq.7 only the UV cut off $\omega_c = m$
must be introduced without IR cutoff in agreement with what occurs
in QED.  If we express in covariant way the quadratic average of velocities
such that the above rest result is reproduced, we must write

\begin{equation}
 \langle{u^{\mu}u^{\nu}}\rangle = - \frac{1}{3}
 \big(1+\frac{\alpha}{\pi}\big)g^{\mu\nu} +
 \frac{\pi^{\mu}}{m} \frac{\pi^{\nu}}{m}\hspace{.4in}.\end{equation}

In order that eq.4 be equal to BMT eq.1, which is necessary to have Thomas
$\frac{1}{2}$ only in normal magnetic moment and not in its anomalous part
in spin orbit interaction, we choose $c^{~}_{3} = 3\frac{e}{2m} - 2\mu'$ so
that

\begin{equation}
\frac{da^{\mu}}{d\tau} = 2\frac{e}{2m} \big(1 + \frac{\alpha}{2\pi} \big)
F^{\mu\nu}a_{\nu} - 2\mu' \frac{\pi^{\mu}}{m} \frac{\pi_{\nu}}{m}
F^{\nu\lambda}a_{\lambda}\hspace{.4in}.
\end{equation}

The difference is now that since we consider instantaneous $u^{\mu}$ and
global $\frac{\pi^{\mu}}{m}$ velocities there is an arbitrariness in
the definition of Pauli-Lubansky vector according to which one we introduce
in eq.2 and it is not obvious that $\pi \cdot a=0$.  If this orthogonality
does not hold, in the rest frame there is a $\xi_0$ apart from
$\boldsymbol{\xi}$  to characterize the polarization.

It is interesting that something similar regarding $\xi_0$ emerges from
the semiclassical limit of Dirac equation~\cite{Rafanelli}.
In fact, taking $\hbar \rightarrow 0$ for Dirac equation for $\Psi$ in
presence of electromagnetic field with the definition

\begin{equation}
\Psi = \Phi R\hspace{.1in} exp \frac{i}{\hbar}S
\end{equation}

\noindent where R and S are real functions and $\Phi$ is a complex 4-spinor
with the conditions

\begin{equation}
\bar{\Phi}\Phi = 1\hspace{.4in},\hspace{.4in} \bar{\Phi} \gamma^{5}\Phi = 0
\end{equation}

\noindent the equation of order $\hbar$ for $\Phi$ shows that with the
identification

\begin{equation}
\frac{\pi^{\mu}}{m} = \bar{\Phi}\gamma^{\mu}\Phi\hspace{.4in},
\hspace{.4in}a^{\mu} = \bar{\Phi}\gamma^{\mu}\gamma_{5}\Phi
\end{equation}

\noindent $\pi^{\mu}$ and $a^{\mu}$ satisfy the Lorentz force and BMT with
$\mu'=0$ respectively.

Now in the global rest frame

\begin{equation}
\xi_{0} = \Phi^{\dagger}\gamma_{5}\Phi
\end{equation}

\noindent is in general non-vanishing in presence of electromagnetic fields
because
Dirac equation couples $p_{\mu}-eA_{\mu}$ to the zitterbewegung of charge
and a non-vanishing weak bispinor of $\Phi$ may survive.  We must note that,
since

\begin{eqnarray}
\gamma_{\mu}\gamma_{5} = - \frac{1}{3!}\epsilon_{\mu\nu\rho\tau}\gamma^{\nu}
\sigma^{\rho\tau}\hspace{.4in}, \nonumber \end{eqnarray}

\noindent this possibility of non-vanishing $\xi_0$ is a consequence of having
defined the Pauli-Lubansky vector with instantaneous velocity and not with
the global one.

An analogy of the Dirac zitterbewegung is given by the classical
Lagrangian~\cite{Barut}

\begin{equation}
L = -i\bar{z}\dot{z} + p_{\mu} (\dot{x}^{\mu} - \bar{z}\gamma^{\mu}z) +
eA_{\mu}\bar{z}\gamma^{\mu}z
\end{equation}

\noindent where $z$ is a 4-spinor.

Defining

\begin{equation}
v^{\mu} \equiv \dot{x}^{\mu} = \bar{z}\gamma^{\mu}z\hspace{.4in},\hspace{.4in}
S^{\mu\nu} = - \frac{i}{4}\bar{z} [\gamma^{\mu},\gamma^{\nu}]z
\end{equation}

\noindent together with $P_{\mu}=p_{\mu}-eA_{\mu}$, the classical equations
of motion

\begin{eqnarray}
\frac{dP_{\mu}}{d\tau} &=& eF_{\mu\nu}v^{\nu}\nonumber\\
\frac{dv_{\mu}}{d\tau} &=& -4S_{\mu\nu}P^{\nu}\nonumber\\
\frac{dS_{\mu\nu}}{d\tau} &=& v_{\mu}P_{\nu} - P_{\mu}v_{\nu}
\end{eqnarray}

\noindent are equal to those obtained for the operators $p_{\mu}-eA_{\mu}$,
$\gamma^{\mu}$ and $[\gamma^{\mu}, \gamma^{\nu}]$ from the Dirac Hamiltonian
with the definition

\begin{equation}
\frac{d}{d\tau}O = -i\gamma^{0}[O,H_{D}]\hspace{.4in}.
\end{equation}

Whereas the contraction

\begin{equation}v \cdot P = m
\end{equation}

\noindent valid for the classical analogy is also a constant as a Dirac
operator, the modulus of the mass velocity $\frac{P^{\mu}}{m}$ and charge
velocity  $v^{\nu}$ are not constants

\begin{equation}
\frac{d}{d\tau}v^{2} \neq 0\hspace{.3in}\hspace{.4in}, \hspace {.4in}
\frac{d}{d\tau}P^{2} \neq 0
\end{equation}

\noindent indicating that both are subject of zitterbewegung.

In the quantum case obviously $\gamma^{\mu}\gamma_{\mu}$ is a constant.
The difference being that 4x4 matrices take into account also negative
energy contributions whereas eq.19 refers to a positive energy particle.

If in the classical analogy we wish to define a spin 4-vector as in eq.2
we have the arbitrariness of using either the mass or the charge velocity.
Choosing the former for the definition of the spin 4-vector and the latter
for its inverse, i.e.

\begin{eqnarray}
S^{\mu} &=&
\epsilon^{\mu\nu\rho\sigma}\frac{P_{\nu}}{m}S_{\rho\sigma}\nonumber\\
S^{\mu\nu} &=& \frac{1}{2}\epsilon^{\mu\nu\rho\sigma}v_{\rho}S_{\sigma}
\end{eqnarray}

\noindent which are compatible, we obtain

\begin{equation}
\frac{d}{d\tau}S_{\mu} =
\frac{e}{m}v_{\mu}v_{\alpha}F^{\alpha\nu}S_{\nu}\hspace{.4in}.
\end{equation}

{}From eq.20 it is by no means evident that the zero component of the
polarization will vanish in the rest frame.

Considering that BMT refers to averages, and assuming that
$\langle v_{\mu}v_{\alpha}S_{\nu}\rangle = \langle v_{\mu}v_{\alpha}\rangle
\langle S _{\nu}\rangle$, eq.21 gives
BMT without anomalous contribution to the magnetic moment if

\begin{equation}
\langle{v_{\mu}v_{\nu}}\rangle = g_{\mu\nu}\hspace{.4in}.
\end{equation}

This average does not seem sensible for spatial indices but this invariant
expression is consistent with $\gamma^{1}\gamma^{1} = - \alpha_{x}\alpha_{x}$
which must be used in quantum Dirac theory.  This shows that BMT corresponds to
averages of quantum quantities and not of classical ones.

Therefore to take radiative effects into account in a semiclassical way,
our approach seems more appropriate.  Instead of considering the detailed
description with one velocity for charge and another for mass, we have an
effective common velocity.
In addition to one Bohr magneton originated from the non coincidence
of charge and mass, another one comes from this common velocity which
affected by the longer wavelength radiation field produces the anomalous
correction.

\section{Effect of large static electromagnetic fields}
\label{Effect of large static electromagnetic fields}

In our approach we must take $a^{\mu}$ as the average of the product of the
common charge-mass instantaneous velocity and spin

\begin{equation}
a^{\mu} = \epsilon^{\mu\nu\rho\sigma} \langle{u_{\nu}S_{\rho\sigma}}\rangle
\hspace{.4in}.\end{equation}

   In the global rest frame with absence of electric and magnetic fields
   $a^{0} = \langle {\bf{u} \cdot \bf{S}} \rangle = 0$.  The same happens
   when these fields are applied together with the spin produced by
zitterbewegung because
of the difference of scale of both motions. But if we consider the
additional angular momentum caused by the precession around the magnetic
field ${\bf{B}} = B{\bf{z}}$ in a range given by the Compton length

\begin{equation}
M_{12} = e\lambda^{2}_{c}B
\end{equation}

\noindent the simultaneous application of ${\bf{E}} = E{\bf{z}}$ will produce
an
asymmetry in the average of the velocity along this axis.  This asymmetry
is given by the electric force times the interval over which the average
must be taken. But this interval decreases with increasing force because
it must correspond to the time in which the electron may be considered
globally at rest.  It is reasonable to take this interval as that necessary
to displace the electron by one Compton length which will be
$\sim {E}^{-1/2}$ for non-relativistic motion and will tend to $\sim{E}^{-1}$
for relativistic one.  Therefore we may state as a bound

\begin{equation}
{\xi}_0 \leq e\lambda^2_c B\hspace{.4in}.
\end{equation}

It must be noticed that for a magnetic field $B \simeq 1T$,
${\xi}_0 \leq 10^{-10}$ which is just below the theoretical error
in the radiative calculations of anomalous gyromagnetic factor.

It is interesting to see which is the correction given by a non-vanishing
${\xi}_0$ to the spin-orbit interaction as described by BMT.
A lengthy but straightforward calculation for an electron with global
momentum $\boldsymbol{\pi}$ under external fields $\bf{E}$ and $\bf{B}$ gives,
as equation of motion for its spin

\begin{eqnarray}
\frac{d{\boldsymbol{\xi}}}{dt} = \frac{2\mu m+2\mu'(\epsilon-m)}{\epsilon}
(\boldsymbol{\xi} \times {\bf B}) + \frac{2\mu'}{\epsilon(\epsilon+m)}
(\boldsymbol{\pi} \cdot{\bf B})(\boldsymbol{\pi} \times
\boldsymbol{\xi})\nonumber\\
+ \frac{2\mu m +2\mu'\epsilon}{\epsilon(\epsilon+m)} \boldsymbol{\xi} \times
({\bf E} \times \boldsymbol{\pi})\nonumber\\ +2\mu'\xi_0 [{\bf E} - \frac{(
\boldsymbol{\pi}\cdot{\bf E}) \boldsymbol{\pi}} {\epsilon(\epsilon+m)} + \frac{
\boldsymbol{\pi}\times{\bf B}}{\epsilon}] \hspace{.3in}.
\end{eqnarray}

The second line of eq.26 gives rise to the spin-orbit interaction which,
in the non-relativistic limit, shows the ``Thomas $\frac{1}{2}$'' for normal
magnetic moment and absent for its anomalous correction.

Thinking that the applied $\bf{B}$ in the electron rest frame may give a
bound for $\xi_0$ as above and for $\frac{\boldsymbol{\pi}}{m} \sim
10^{-2}$, the ratio between the third line to the second one is
$\leq10^{-11}$.  Since spin-orbit interaction produces a fine structure
$\sim 10^{-3}eV$ the effect of ${\xi}_0$, which is of electric
dipole type proportional to the anomalous part of the magnetic moment, is
$\leq10^{-14}eV$ much smaller than the experimental error in the Lamb shift
$\sim10^{-11}eV$.

Obviously it is conceivable that an application to an atom is beyond the
validity range of the semiclassical approximation.

Another possible effect of ${\xi}_0$ is on the helicity change of a
free electron under the influence of external uniform e.m. fields.  In this
case the semiclassical approximation should be valid up to
high~\cite{Baier1} magnetic fields $B < \frac{m^{2}}{e} \sim 10^{9} T$ and
even higher for very relativistic electrons.  Since

\begin{equation}
\frac{d\xi_\parallel}{d\tau} = \frac{1}{m} (a_{\parallel}
\frac{d\epsilon}{d\tau} - a_{0}\frac{d\mid \boldsymbol{\pi}\mid}{d\tau} +
\frac{da_{\parallel}}{d\tau}\epsilon -
\frac{da_{0}}{d\tau}\mid{\boldsymbol{\pi}} \mid)\hspace{.4in}, \end{equation}

\noindent it is easy to see that the second term of eq.9 being proportional to
$\pi^{\mu}$ gives no contribution to eq.27.  On the other hand, with
longitudinal fields, whereas the first two terms in brackets of eq.27 give
$eF^{03}\xi_0$, the last two terms contribute as
$-e(1+\frac{\alpha}{\pi})F^{03}\xi_0$.  Therefore

\begin{equation}
\frac{d\xi_\parallel} {d\tau} = \frac{e}{m} \frac{\alpha}{2\pi}
E_{\parallel}\xi_0
\end{equation}

\noindent and taking the volume
$V = {\sqrt{1-\frac{\pi^2}{\epsilon^2}} \lambda^{3}_{c}}$
for each spin

\begin{equation}
\frac{1}{V}\frac{d\xi_\parallel}{dt} \leq \frac{\alpha}{2\pi}e^{2}
E_{\parallel}B_{\parallel}\hspace{.4in}.
\end{equation}

This may be understood as the effect of the chiral anomaly~\cite{Nielsen}
on the helicity of the electron in a way analogous to the Feynman graphs
analysis of ref.~\cite{Teryaev2}.

One must remark that $\xi_0 \neq 0$ has the consequence that from eq.9
it does not follow that $\frac{da^{2}}{d\tau} = 0$, and also in general
$\frac{d}{d\tau} \mid {\boldsymbol{\xi}}\mid^2 \neq 0$.  Compared to the Dirac
equation, in the Foldy-Wouthuysen approach $\mid{\boldsymbol{\xi}}\mid^2$ is
constant, but this does not happen in the 4-spinor treatment where only the
4x4 matrix $\mid{\boldsymbol{\Sigma}}\mid^2$ is a constant.  The fact that our
result gives a non constancy of $\mid{\boldsymbol{\xi}}\mid^2$ should be due to
the inclusion in the one particle description of sea effects produced by
strong external fieds.

{}From eq.29 for an electron traveling 1 cm at not too high velocity under
longitudinal $E \approx B \sim 10^{2}(eV)^{2} \simeq 1T$ the change of
helicity would be $\leq10^{-12}$, being necessary to take into account that
for ultrarelativistic electron $\Delta\xi_{\parallel} \rightarrow 0$
so that the helicity cannot increase indefinitely.

We finally comment that eq.29 can also be intuitively understood as the
influence of the anomalous part of the magnetic moment which allows a
different precession of instantaneous velocity and spin around the magnetic
field so that taking the average, the longitudinal electric field produces
an asymmetry and a net variation of helicity results.

\section{Acknowledgments}
          \label{Acknowledgments}

A.D.S. thanks the hospitality at the Centro At\'omico Bariloche and
L.M. that at the Dipartimento di Scienze Fisiche of Naples during parts of
this work.

This research was partially supported by CONICET Grant No. 3965/92.

\end{document}